\documentclass[aps,pre, twocolumn]{revtex4-1}
\usepackage{amsmath,amsfonts,bm, color,amssymb}
\usepackage{graphicx,epsfig,overpic,hyperref,ulem}
\usepackage{subfigure,float}

\newcommand{\refeq}[1]{Eq. \ref{#1}}
\newcommand{\reffig}[1]{FIG. \ref{#1}}

\newcommand{\eps}{{\varepsilon}}
\newcommand{\bE}{{\bf E}}
\newcommand{\bF}{{\bf F}}

\newcommand{\bT}{{\bf T}}
\newcommand{\bI}{{\bf I}}

\newcommand{\bn}{{\bf n}}

\newcommand{\yhat}{{\bf \hat y}}
\newcommand{\zhat}{{\bf \hat z}}

\newcommand{\bnabla}{{\bm \nabla}}

\begin{document}
\title{Particle-surface interactions in a uniform electric field} 

\author{Zhanwen Wang$^{a}$, Michael J. Miksis$^{b}$ and Petia M.Vlahovska$^{b}$}
\affiliation{%
$^{a}$~Theoretical and Applied Mechanics Program, Northwestern University, Evanston, IL 60208, USA\\
$^{b}$~Engineering Sciences and Applied Mathematics, Northwestern University, Evanston, IL 60208, USA. E-mail: petia.vlahovska@northwestern.edu\\
 }

\date{\today}

\begin{abstract}

The electrostatic force on  a spherical particle near a planar surface is calculated for the cases of a uniform electric field  applied in either normal or tangential direction to the surface. The particle and suspending media are assumed to be weakly conducting, so that that the leaky dielectric model applies. The Laplace equation for the  electric potential is solved in bipolar coordinate system and the potential is obtained in terms of a series expansion of Legendre polynomials. The force  on the particle is calculated using the Maxwell tensor.  We find that in the case of normal electric field, which corresponds to a particle near an electrode, the force is always attractive but at a given separation it varies nontrivially with particle-suspending medium conductivity ratio; the force on a particle that is more conducting than the suspending medium is much larger compared to the force on a particle less conducing than the suspending medium.
 In the case of tangential electric field, which  corresponds to a particle near an insulating boundary, the force is always repulsive. 



\end{abstract}

\maketitle

\section{Introduction}
Electric fields are a classic means to manipulate and assemble colloidal particles \cite{Motosuke2017,PRIEVE:2010,Velev:2006,Velev_review:2015}. More recently, electric fields have become a popular tool to energize and create self-propelled particles \cite{Yan:2016,Han:2018,Driscoll:2019} due to field-induced charge electrophoresis \cite{Velev:2008, Ma-Wu:2015, Nishiguchi:2015} or torque (due to the Quincke effect), which drives colloids to roll on a surface  \cite{Bartolo:2013, Bartolo:2015,Karani:2019,Gerardo:2019,Zhang:2021a}.  In these applications particles are in close proximity to boundaries, and the electrostatic force (and torque ) exerted on the particle  is significantly influenced by confinement.
The canonical problem of a particle near an electrode in the presence of a uniform electric field has been analyzed in the literature theoretically in the two limiting cases of a conducting or an insulating sphere. 
The surface of a conducting particle is equipotential, and consequently, the electric field inside vanishes. The net charge and force on a spherical particle are found using  the method of images \cite{Perez:2002,Bishop:2014} or the equivalent problem of two spheres in a uniform electric field \cite{Davis:1964}. If the particle is perfect dielectric, the boundary condition on the particle-medium interface are  continuity of the electric potential and a jump of the displacement field due to given surface charge. The electrostatic force  has been found either in terms of series expansion in eigenfunctions 
of the Laplace equation in bispherical coordinate system \cite{Lin:1994,Cai:1987,Danov:2006,Stoy:1989b} or from a multipole-moment theory for the pair-wise  dielectrophoretic interactions of dielectric spheres \cite{Washizu:1996}.

In this paper, we solve for the electrostatic force on a particle with arbitrary conductivity. In this case,  the boundary conditions on the particle interface are continuity of the electric potential and the normal electric current, and the bulk media is assumed to be charge free. The surface charge density needed to satisfy the charge flux continuity is thus determined as part of the problem. This is  the so called ``leaky-dielectric'' model \cite{Taylor:1966}.  Using this approach, the force on a sphere straddling a planar interface subjected to either a normal or tangential uniform electric field has been calculated \cite{Yi:2019}. However, the force on a sphere near an interface has not be derived, even though the electric field for the equivalent problem of two leaky-dielectric spheres with line-of-centers parallel or perpendicular to the applied field has been derived \cite{Stoy:1989b, stoy1989solution}. 
This paper is organized as follows. The problem is formulated in Section 2. In Section 3, the solution methodology using bispherical coordinates is presented. First, the solution of the electric field is expressed in terms of a series of Legendre polynomials, and an algorithm to determine its coefficients is given. Then, the electric field strength, surface charge distribution on the interface and the force on the particle are calculated. Details are presented in Appendix. In Section 4, we validate the solution with the published results for the interaction force between two identical dielectric spheres in a uniform electric field \cite{Washizu:1996} and  a conducting sphere near an electrode \cite{Bishop:2014,Perez:2002}, and study the force dependence on separation and particle electric properties. 


\section{Problem formulation}

\begin{figure}[h]
	\centering
	\subfigure[]{\includegraphics[width=0.6\linewidth]{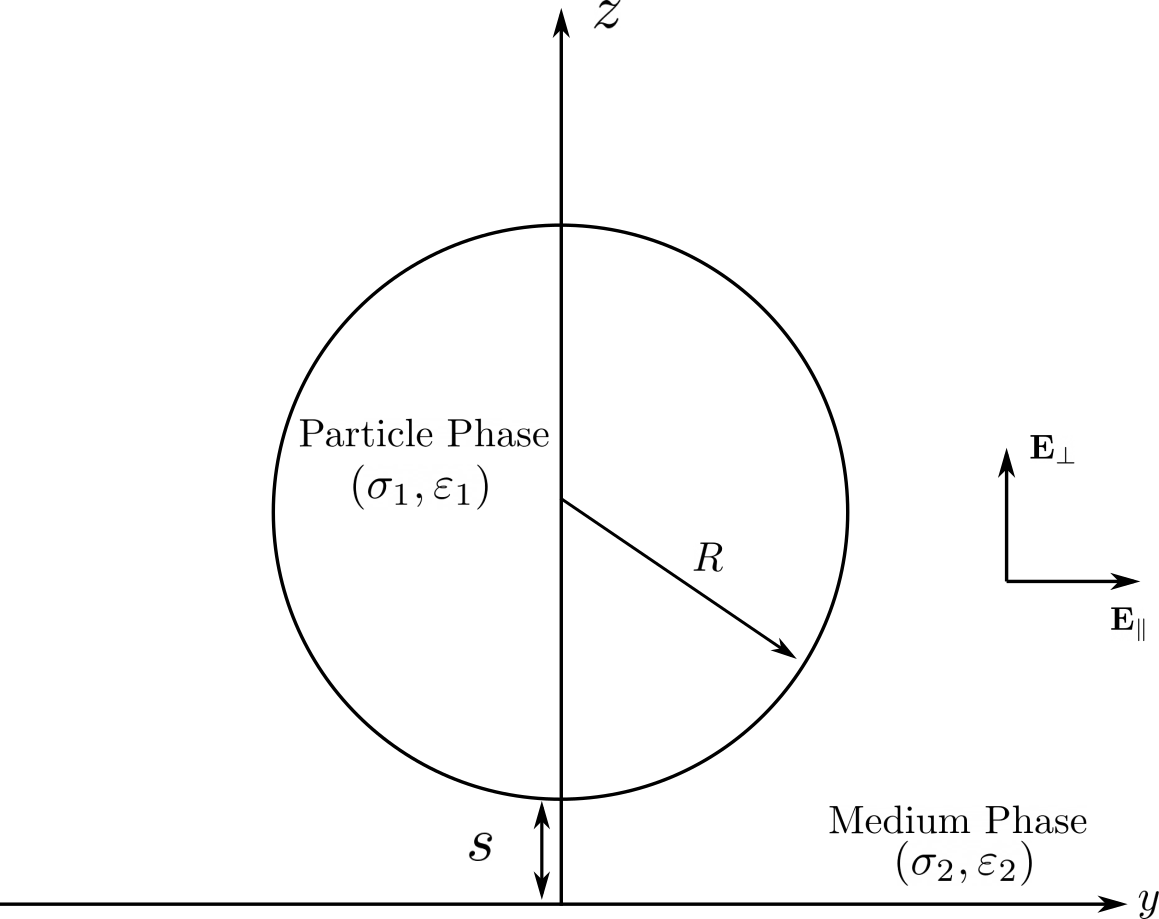}}
	\subfigure[]{\includegraphics[width=0.6\linewidth]{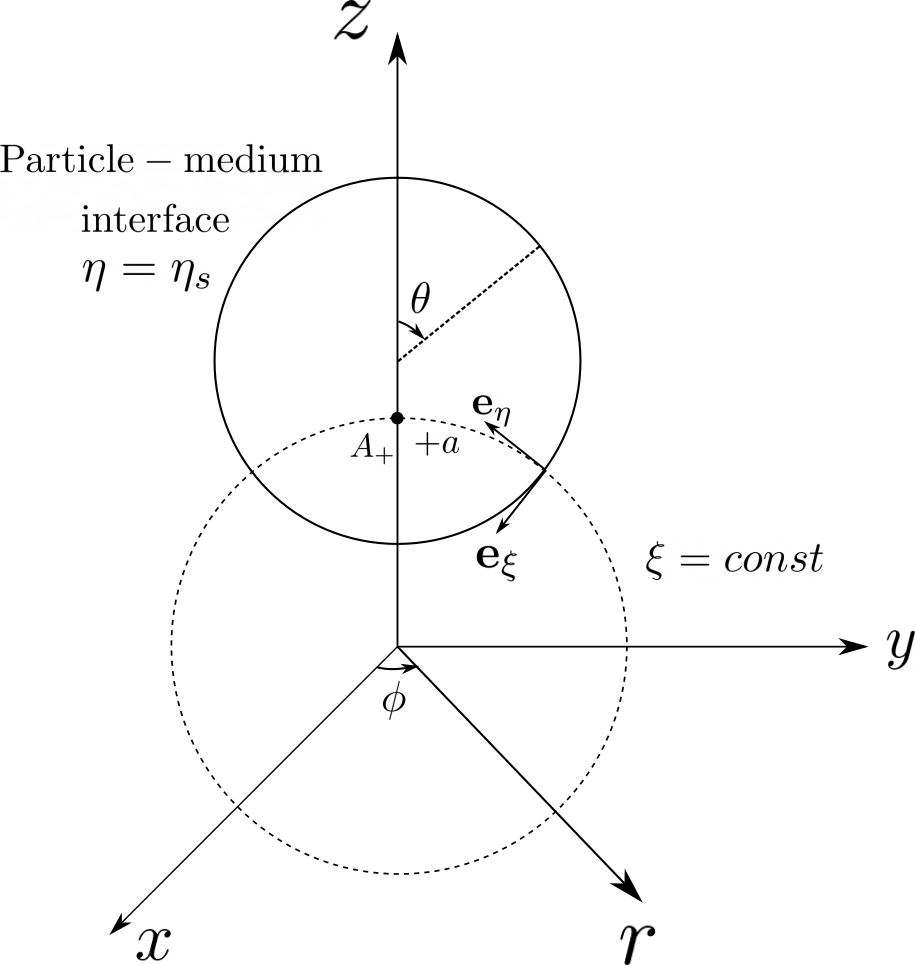}}
	\caption{(a) A spherical particle with radius $R$ is suspended above a planar wall.  The gap between the particle and wall surfaces is $s$. The applied electric field is either in the normal direction $\mathbf{E}_{\perp}=E_0\zhat$ with a conducting wall or in the tangential direction $\mathbf{E}_{||}=E_0\yhat$ with an insulating wall. 
	(b) Sketch of Cartesian coordinates $\left(x,y,z\right)$, cylindrical coordinates $\left(r, z, \phi \right)$ and bipolar coordinates $\left( \xi, \eta, \phi \right)$ with unit coordinate vectors $\boldsymbol{e}_{\xi}$ and $\boldsymbol{e}_{\eta}$, and a pole $A_+$. The isosurfaces of coordinates $\eta$ and $\xi$ are obtained by rotating about $z$ axis. }
	\label{fig:sketch}
\end{figure}

Consider a spherical particle with radius $R$, conductivity $\sigma_1$, and permittivity $\eps_1$ suspended in a medium with conductivity $\sigma_2$ and permittivity $\eps_2$. The particle is  in the vicinity of a boundary and the surface-to-surface distance is $s$.  A uniform electric field is applied in a direction either normal or tangential to the boundary.
The problem is sketched in  \reffig{fig:sketch}a.



We adopt the leaky dielectric model, which assumes an irrotational electric field ($\bE=-\nabla\Phi$, where $\Phi$ is the electric potential), electroneutral bulk fluids, and that electric current obeys Ohm’s law
\cite{Melcher-Taylor:1969}. Accordingly, the electric potential inside the particle, $\Phi_{1}$, and in the suspending medium, $\Phi_{2}$, satisfy the Laplace equation
\begin{equation}\label{governing_equation}
	\nabla^{2}\Phi_{1}=0\,,\quad \nabla^{2}\Phi_{2}=0 .
\end{equation}
At the particle-medium interface,  the potential and normal electric current are continuous 
\begin{equation}\label{bc2}
	\Phi_{1}=\Phi_{2}\,,\quad \sigma_{1}\bn\cdot \bE_1= \sigma_{2} \bn\cdot\bE_2,
\end{equation}
where $\bn$ is the inward normal to the interface. The current leads to accumulation of charge $Q$ at the surface, which is calculated a posteriori from the jump in the electric displacement field at the interface,
$$
Q = \left(\eps_1\bE_1 - \eps_2\bE_2\right)\cdot \bn.
$$
Away from the particle, the electric field is undisturbed.
We consider two cases: (i) an electric field applied in a direction normal to a boundary that is an equipotential surface, $\bE_\perp=E_0\zhat$ at infinity, and  $\Phi_2=0$ at $z=0$, and (ii) an electric field applied in a direction tangential to an insulating boundary, $\bE_{||}=E_0\yhat$, and  $\bn\cdot \bE_2=0$ at $z=0$.

We introduce disturbance fields $\hat{\Phi}_{1}$ and $\hat{\Phi}_{2}$,
\begin{equation} \label{disturbance_field}
	\hat{\Phi}_{1}=\Phi_{1}+E_0 x_\alpha,~\hat{\Phi}_{2}=\Phi_{2}+E_0 x_\alpha,
\end{equation}
where $x_\alpha$ is the direction of the externally applied electric field. In the case of normal electric field, $x_\alpha = z$, and in the case of tangential electric field, $x_\alpha=y$. The governing equation becomes
\begin{equation}
	\nabla^{2}\hat{\Phi}_{1}=0,~\nabla^{2}\hat{\Phi}_{2}=0 \label{eq_d}.
\end{equation}
The boundary condition at infinity is now homogeneous, $\hat{\Phi}_{2}\to 0$.
At the planar boundary, in the case of a normal electric field, 
\begin{equation}
\hat\Phi_2=0~{\rm at}~z= 0\quad  \mbox{(normal electric field).}
	\label{BoundaryConditionBottomZ}
\end{equation}
or,  in the case of a tangential electric field,
\begin{equation}
\frac{\partial \hat{\Phi}_2}{\partial z}=0~{\rm at}~z= 0\quad  \mbox{(tangential electric field).}		
	\label{BoundaryConditionBottomY}
\end{equation}
The boundary conditions on the particle-medium interface become
\begin{equation}
	\hat{\Phi}_{1}=\hat{\Phi}_{2}, \label{ElectricPotentialContinuity}
\end{equation}
and
\begin{equation}
\label{CurrentContinuity}
	\sigma_{1}\frac{\partial\hat{\Phi}_{1}}{\partial n} - \sigma_{2}\frac{\partial\hat{\Phi}_{2}}{\partial n} = E_{0}\left(\sigma_{1}-\sigma_{2}\right)\frac{\partial x_{\alpha}}{\partial n},
\end{equation}
The problem is solved using separation of variables in bispherical coordinates  $\left( \xi, \eta, \phi \right)$\cite{Danov:2006}, defined as 
\begin{equation}\label{cy_to_bsp}
	r=\frac{a}{h}\sin \xi,~z=\frac{a}{h}\sinh \eta,~h\equiv\cosh\eta-\cos\xi.
\end{equation}
where $a = \left[ (2R+s) s \right]^{1/2}$.
\reffig{fig:sketch}b illustrates the  coordinate systems  that are used in this analysis. 
All isocurves of coordinate $\xi$ pass through poles $A_+$ and $A_-$, $A_{\pm} = (0,0,\pm a)$ in Cartesian coordinates $(x,y,z)$, as $\eta \to \pm \infty$.
The particle-medium interface is specified  by $\eta = \eta_s$
\begin{align}
	&\eta_s = \ln \left\{ 1 + \frac{s}{R} + \left[ \left( 2+\frac{s}{R} \right) \frac{s}{R} \right] ^ {1/2} \right\} \label{eta_s}.
\end{align}
The inward normal to the particle surface is $\mathbf{e}_{\eta}$. Hence, \refeq{ElectricPotentialContinuity} and \ref{CurrentContinuity} are written as
\begin{gather}
\hat{\Phi}_1 = \hat{\Phi}_2,~{\rm at}~\eta = \eta_s \label{PotentialContinuityRewritten} \\
\sigma_{1}\frac{\partial\hat{\Phi}_{1}}{\partial \eta} - \sigma_{2}\frac{\partial\hat{\Phi}_{2}}{\partial \eta} = E_{0} \left(\sigma_{1}-\sigma_{2}\right) \frac{\partial x_{\alpha}}{\partial \eta},~{\rm at}~\eta = \eta_s.
\label{CurrentContinuityRewritten}
\end{gather}
The planar wall boundary is specified by $\eta=0$. Accordingly,
\begin{align}
&\hat\Phi_2=0,~{\rm at}~\eta = 0\quad  \mbox{(normal electric field),} \label{BoundaryConditionBottomN}\\
&\frac{\partial \hat{\Phi}_2}{\partial \eta}=0,~{\rm at}~\eta = 0\quad  \mbox{(tangential electric field).}
\label{BoundaryConditionBottomT}
\end{align}
In bispherical coordinates, we require an additional boundary condition, that is, $\hat{\Phi}_{1}$ to be finite at pole $A^+$, where $\eta \to \infty$,
\begin{equation}
	|\hat{\Phi}_{1} |<\infty,~\textup{as}~\eta\to\infty. \label{BCetaInf}
\end{equation}

Finally, note that $r\to \infty$ corresponds to $\eta\to 0$ and $\cos\xi\to 1$. Hence, from \refeq{BoundaryConditionBottomN}, we see that
the boundary condition at infinity is satisfied automatically in the case of normal electric field, while for the tangential electric field, an auxiliary boundary condition is needed,
\begin{equation}
	\hat{\Phi}_2 \sim 0,~{\rm as}~\eta\to 0~{\rm and}~\cos\xi\to 1.
	\label{BCinfRewrittenT}
\end{equation}


\section{Electric potential}
In bispherical coordinates, the Laplace's equation can be written as
\begin{align}
	\nabla^2 F = &\frac{h^3}{a^2 \sin\xi} \left[\sin\xi\frac{\partial}{\partial\eta} \left(\frac{1}{h}\frac{\partial F}{\partial\eta}\right) \right. \nonumber\\
	&\left.+ \frac{\partial}{\partial \xi}\left(\frac{\sin\xi}{h} \frac{\partial F}{\partial\xi}\right)\right] + \frac{h^2}{a^2\sin^2\xi} \frac{\partial^2 F}{\partial\phi^2}=0, \nonumber
\end{align}
where $F$ stands for either $\hat{\Phi}_1$ or $\hat{\Phi}_2$.  The general solution is
\begin{widetext}
\begin{equation}
	{F} = {\sqrt{h}}\sum_{m,n} \left[X_n^m e^{\lambda_n\eta} + Y^m_n e^{-\lambda_n\eta} \right] P_n^m(\cos\xi)\cos(m\phi) \nonumber + \sum_{m,n} \left[M_n^m e^{\lambda_n\eta} + N^m_n e^{-\lambda_n\eta}\right]P_n^m(\cos\xi)\sin(m\phi)\,,
\end{equation}
\end{widetext}
where $\lambda_n \equiv n+1/2$ and $P_n^m$ stands for associated Legendre polynomials of degree $n$ and order $m$. $\sum_{m,n}$ denotes a double sum $\sum_{m=0}^{\infty}\sum_{n=m}^{\infty}$. 

In the case of a uniform electric field, the solution is greatly simplified. For a normal electric field,  the axial symmetry of the problem leads to
\begin{equation}
\label{gsN}
	F = \sqrt{h} \sum_{n=0}^{\infty} \left[X_n^0e^{\lambda_n\eta}+Y^0_n e^{-\lambda_n\eta}\right] P_n(\cos\xi).
\end{equation}
The tangential electric field indicates the dependence on $\phi$ is $\sin\phi$. In this case, the general solution is
\begin{equation}
\label{gsT}
	F = \sqrt{h} \sum_{n=1}^{\infty} \left[M_n^1e^{\lambda_n\eta}+N^1_n e^{-\lambda_n\eta}\right] P_n^1(\cos\xi)\sin\phi.
\end{equation}

From the general solutions \refeq{gsN} and \refeq{gsT}, applying the boundary conditions at the wall and at the particle interface for continuous potential  leads to   
\begin{align}
	\Phi_{1} &= -E_0 x_\alpha + \sqrt{h} \sum_{n=n_0}^{\infty} A_{n} \left[ e^{2\lambda_{n}\eta_{s} } \mp 1 \right] e^{ -\lambda_{n}\eta} {\cal{P}}_{n}, \label{solution1}\\
	\Phi_{2} &= -E_0 x_\alpha + \sqrt{h} \sum_{n=n_0}^{\infty} A_{n} \left[ e^{ \lambda_{n}\eta } \mp e^{-\lambda_{n}\eta} \right] {\cal{P}}_{n} . \label{solution2}
\end{align}
where  ${\cal{P}}_n \left( \cos\xi \right)$ stands for $P_{n} \left( \cos\xi \right)$ and $P^1_n(\cos \xi) \sin\phi$ for   the applied normal and tangential electric fields, respectively. The minus signs in the second term in \refeq{solution1} and \refeq{solution2} are for normal electric field. Note that $P_n^1(1)=0$. Accordingly, \refeq{BCinfRewrittenT} is automatically satisfied. The sum starts at $n_0=0$ and 1 for normal and tangential electric fields, respectively.
Plugging \refeq{solution1} and \refeq{solution2} into the boundary condition \refeq{CurrentContinuityRewritten}, we get the following recurrence formula of $A_{n}$ after some algebra
\begin{equation} \label{ThreeDiagonal}
	L_{n,1} A_{n-1} + L_{n,2} A_{n} + L_{n,3} A_{n+1} = G_n.
\end{equation}
Details are listed in Appendix B.

\begin{figure}[h]
	\centering
	\subfigure[Normal electric field]{\includegraphics[width=\linewidth]{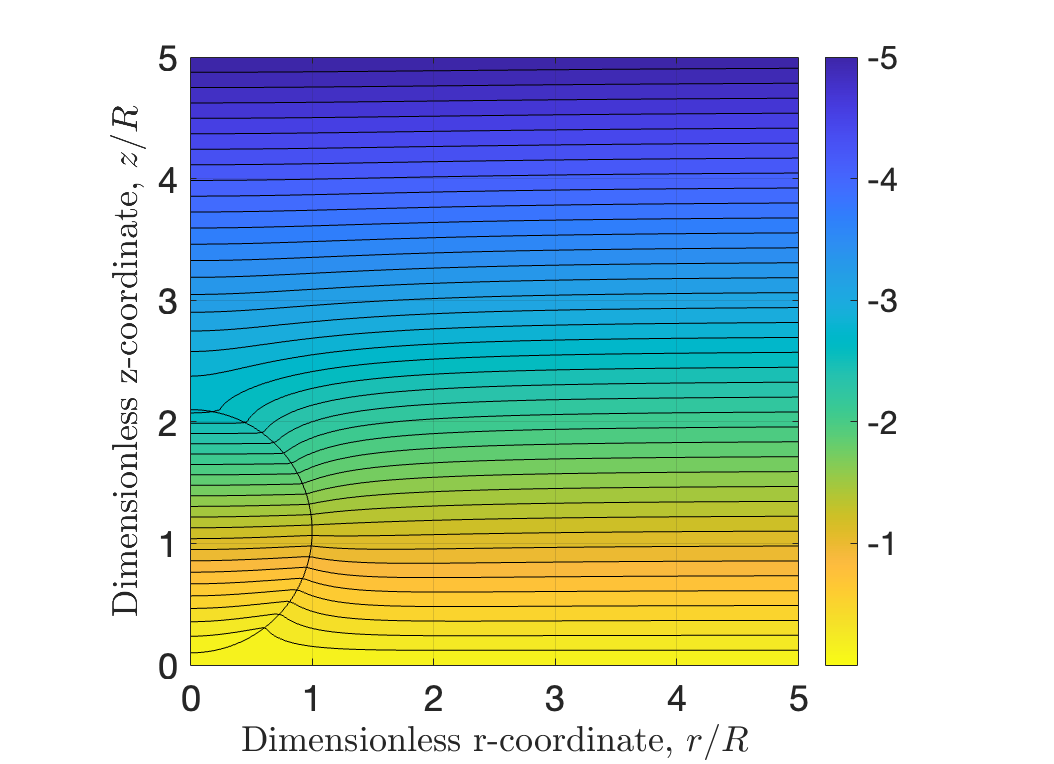}}
	\subfigure[Tangential electric field]{\includegraphics[width=\linewidth]{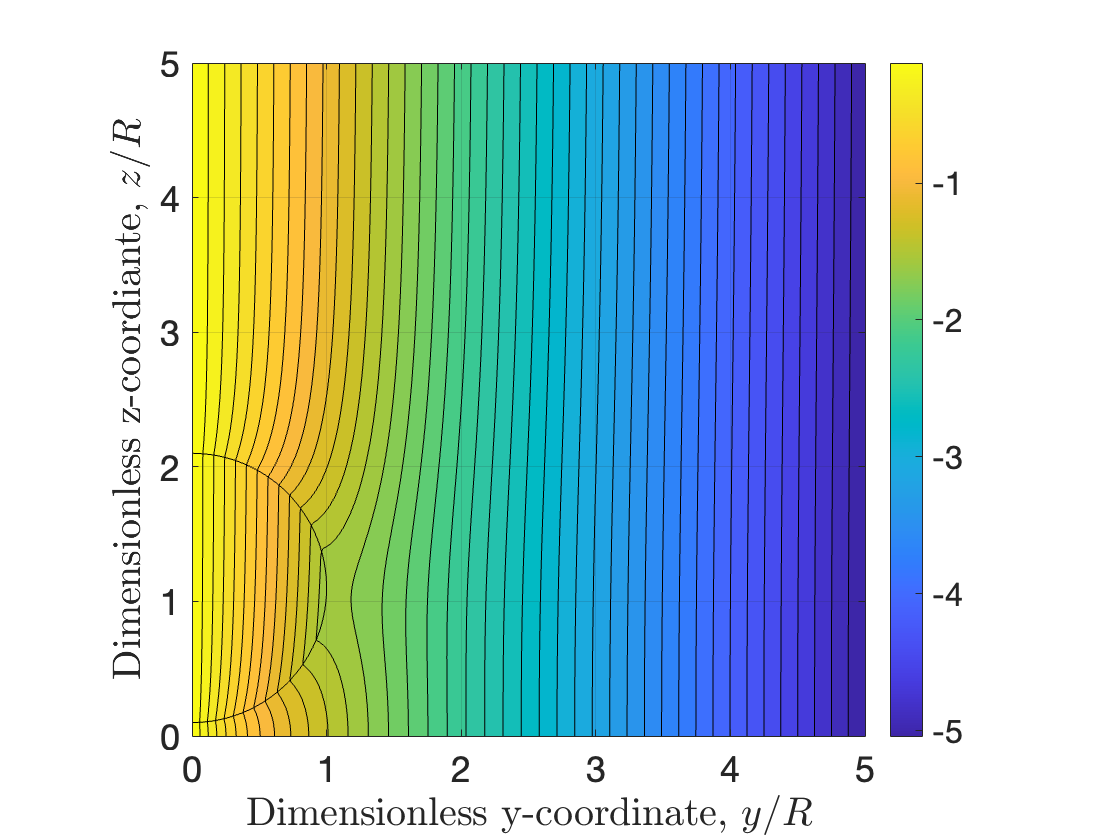}}
	\caption{Equipotential contours for particle-wall separation $s/R=0.1$ and perfectly insulating particle $\beta_{12}=-1$. For the tangential electric field, the equipotential contours are on $x=0$ plane. ($N=50$).}
	\label{fig:equipotential}
\end{figure}
\refeq{ThreeDiagonal} is solved numerically as follows.
 First, we choose a sufficiently large number, $N$, which set the number of terms retained in the series solution. For $n=N$, the last equation in the system \ref{ThreeDiagonal} is
\begin{equation} \label{last}
	L_{N-1,1}A_{N-1}+L_{N,2}A_{N}=G_{N} -  L_{N,3}A_{N+1}.
\end{equation}
where $A_{N+1}$ comes from the asymptotic behavior \refeq{AnAsymptoticN} and \refeq{AnAsymptoticT}. Then, coefficients $A_n~(n \leq N)$ are found from the three-diagonal system (\refeq{ThreeDiagonal} ), whose matrix form is presented below,
\begin{equation}
	[{\bf L}]\cdot {\bf A} = {\bf G},
\end{equation}
where the left-hand-side coefficient matrix $[\mathbf{L}]$ is a triple-diagonal matrix, ${\bf A}$ is the column vector of coefficients $A_n$, and ${\bf G}$ is the column vector of the right-hand-side terms. 

For the normal applied electric field, their components are listed below,
$$
[{\bf L}]
=
\begin{bmatrix}
	L_{0,2} & L_{0,3}  \\
	L_{1,1} & L_{1,2} & L_{1,3} \\
	& L_{2,1} & L_{2,2} & L_{2,3} \\
	&         & \ddots & \ddots & \ddots \\
	&         &        &L_{N-1,1} & L_{N-1,2} & L_{N-1,3} \\
	&         &        &        & L_{N,1}& L_{N,2}
\end{bmatrix}
$$
$$
{\bf A}=
\begin{bmatrix}
	A_0 \\
	A_1 \\
	A_2 \\
	\vdots \\
	A_{N-1} \\
	A_N
\end{bmatrix}
,
{\bf G}=
\begin{bmatrix}
	G_{0} \\
	G_{1} \\
	G_{2} \\
	\vdots \\
	G_{N-1} \\
	G_{N} -  L_{N,3}A_{N+1}
\end{bmatrix}
.
$$
The expressions for $L$, $G$, and $A_{N+1}$ are given by  \refeq{LnN}, \refeq{GnN}, and \refeq{AnAsymptoticN}. 

For the tangential electric field, subscripts start from 1,
$$
[{\bf L}]
=
\begin{bmatrix}
	L_{1,2} & L_{1,3}  \\
	L_{2,1} & L_{2,2} & L_{2,3} \\
	& L_{3,1} & L_{3,2} & L_{3,3} \\
	&         & \ddots & \ddots & \ddots \\
	&         &        &L_{N-1,1} & L_{N-1,2} & L_{N-1,3} \\
	&         &        &        & L_{N,1}& L_{N,2}
\end{bmatrix}
$$
$$
{\bf A}=
\begin{bmatrix}
	A_1 \\
	A_2 \\
	A_3 \\
	\vdots \\
	A_{N-1} \\
	A_N
\end{bmatrix}
,
{\bf G}=
\begin{bmatrix}
	G_{1} \\
	G_{2} \\
	G_{3} \\
	\vdots \\
	G_{N-1} \\
	G_{N} -  L_{N,3}A_{N+1}
\end{bmatrix}
.
$$
The expressions for $L$, $G$, and $A_{N+1}$ are given by \refeq{LnT} - \ref{AnAsymptoticT}. 

The solution appears spectrally convergent at sufficiently high $N$. As $N$ increases, we observe that the error decreases as $1/N^2$ and approaches near plateau as $N$ increases  \cite{Zhanwen:thesis}. We find that  the number of terms $N$ needed for the force coefficient $C_f$ to reach $10^{-5}$ relative error,
$$
\frac{\left| C_f\left(N+1\right) - C_f\left(N\right) \right|}{\left|C_f\left(N\right)\right|} \leq 10^5,
$$
scales as $10\delta^{-1/2}$, independently of  the conductivity and permitivity ratios.
We validated our solution for a conducting particle with the result
derived using electrostatic images  \cite{Perez:2002}, see 
 \reffig{fig:cfvsdelta}. We also compared with the two-sphere solution of \cite{Washizu:1996}. In both cases, the agreement is excellent.



\section{Electric field and induced surface charge distribution}
The electric field is obtained by calculating the gradient of the electric potential,
\begin{equation} \label{elefield}
	\bE_{i}=-\bnabla \Phi_{i} = E_{i,\eta} {\bf e}_{\eta} + E_{i,\xi} {\bf e}_{\xi} + E_{i,\phi} {\bf e}_{\phi},
\end{equation}
where the subscript $i$ is the index of the phases. $\bE_1$ and $\bE_2$ are the electric field strengths inside and outside the particle, respectively.
Substituting \refeq{disturbance_field} into \refeq{elefield} yields the component form,
\begin{align}
	E_{i,\eta} &= -\frac{h}{a} \left( \frac{\partial \hat{\Phi}_{i}}{\partial \eta} - E_{0} \frac{\partial x_{\alpha}}{\partial \eta} \right), \label{Ei_eta}\\
	E_{i,\xi} &= -\frac{h}{a} \left( \frac{\partial \hat{\Phi}_{i}}{\partial \xi} - E_{0} \frac{\partial x_{\alpha}}{\partial \xi} \right),  \label{Ei_xi}\\
	E_{i,\phi} &= -\frac{h}{a\sin\xi} \left( \frac{\partial \hat{\Phi}_{i}}{\partial \phi} - E_{0} \frac{\partial x_{\alpha}}{\partial \phi} \right), \label{Ei_phi}
\end{align}

The discontinuity of the electric field across the particle-medium interface leads to accumulation of  charge $Q$. The surface density of this induced  charge is calculated from
\begin{equation} \label{SurfaceChargeDistribution}
	Q=\left. \left( \varepsilon_1 E_{1,\eta} - \varepsilon_2 E_{2,\eta}\right) \right|_{\eta = \eta_{s}}.
\end{equation}

In the case a normal applied field, due to the symmetry of the problem, $E_{i,\phi} = 0$. Explicit expressions for $E_{i,\eta}$ and $E_{i,\xi}$ are listed in Appendix A. For a net charge-neutral particle, we find the following identity satisfied by the coefficients $A_n$
\begin{equation} \label{AnSummation}
	\sum_{n=0}^{\infty} A_n = 0,
\end{equation}
see Appendix C for details.

In the case of the tangential applied  electric field, the electric field components are given by
\begin{align}
	E_{i,\eta} &= \hat{E}_{i,\eta} \sin\phi, \nonumber \\
	E_{i,\xi} &= \hat{E}_{i,\xi} \sin\phi, \nonumber \\
	E_{i,\phi} &= \hat{E}_{i,\phi} \cos\phi, \nonumber
\end{align}
and $\hat{E}_{i,\eta}$, $\hat{E}_{i,\xi}$, and $\hat{E}_{i,\phi}$ are listed in Appendix A. \refeq{SurfaceChargeDistribution} shows that $Q$ for a tangential electric field has the sine dependence of $\phi$,
\begin{equation}
	Q=\hat{Q}\sin\phi. \label{SurfaceChargeT}
\end{equation}


\begin{figure}[h]
	\centering
	\subfigure[Normal electric field]{\includegraphics[width=\linewidth]{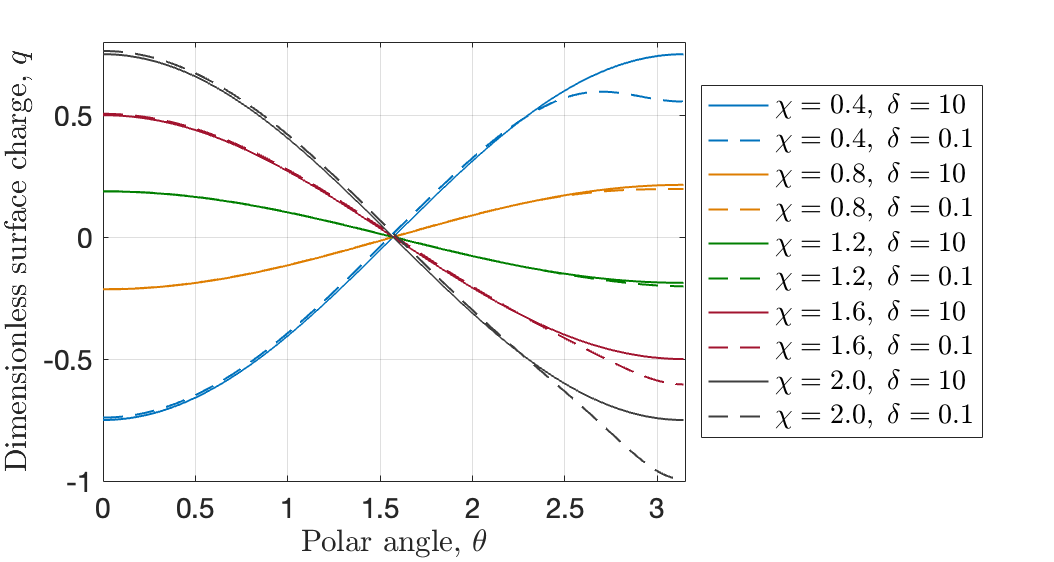}}
	\subfigure[Tangential electric field]{\includegraphics[width=\linewidth]{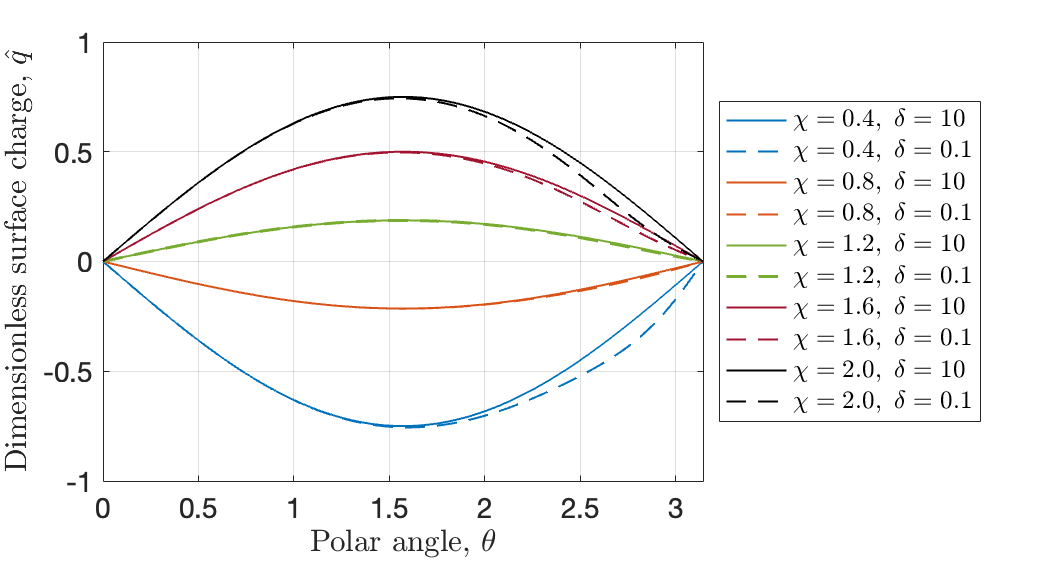}}
	\caption{Dimensionless surface charge density, $q=Q/(\varepsilon_2 E_0)$, as a function of the polar angle  $\theta$ for various values of the conductivity ratio $\chi=\sigma_1/\sigma_2$ and particle-wall separations $\delta=s/R$.  In the case of the tangential applied field, $q = \hat{q}\sin\phi$ and the $\theta$ dependence of $\hat q$ is shown. $\theta$ is the angle between the positive $z$-axis and the line from the particle center to a point on the interface, see \reffig{fig:sketch}b. Permittivity ratio $\kappa=\eps_1/\eps_2=1$.
	}
	\label{fig:surfacecharge}
\end{figure}

\reffig{fig:surfacecharge} illustrates the dependence of the dimensionless surface charge distribution, $q=Q/(\varepsilon_2 E_0)$, on dimensionless particle-wall separation $\delta = s/R$ and conductivity ratio $\chi=\sigma_1/\sigma_2$ with permittivity ratio $\kappa=\eps_1/\eps_2=1$. 
If $\kappa > \chi$, i.e., $\varepsilon_1/\sigma_1>\varepsilon_2/\sigma_2$, the surface charge distribution indicates a free-charge dipole which is antiparallel to the applied  electric field.  In the opposite case, $\kappa < \chi$, 
the dipole is in the same direction as the applied  electric field. At large separations ($s/R=10$), the distribution is symmetric. The  wall proximity introduces asymmetry in the charge distribution. However,   the surface integral over the  particle-medium interface (a unit sphere in dimensionless form),
is  zero and the particle is net charge neutral.


\section{Interaction force}
The electrostatic force on the particle is calculated from the Maxwell stress
\begin{equation} \label{Force}
	\bF = \int_{S_p} \left(\bT_{1}-\bT_{2}\right)\cdot\mathbf{e}_{\eta}  dS\,,
\end{equation}
where $\mathbf{e}_{\eta}$ is the inward normal on the interface, and $\bT_{1}$ and $\bT_{2}$ are Maxwell stress tensor in the particle and media phases, respectively. 
\begin{equation}\label{MaxwellTensor}
	\bT_{i} = \eps_i \left[\bE_i\bE_i - \frac{1}{2}\left(\bE_i\cdot\bE_i\right)\bI\right] \,.
\end{equation}
The divergence of the Maxwell strength is zero, $\bnabla \cdot\bT_{i}=0$, because of the bulk media are charge-free. Using the divergence theorem leads to
\begin{equation}
	\bF = -\int_{S_B} \bT_2^D \cdot \mathbf{e}_{\eta} dS,
	\nonumber
\end{equation}
where $S_B$ stands for the wall boundary and $\mathbf{T}_2^D$ is the disturbance Maxwell stress tensor,
\begin{equation}
\begin{aligned}
	\bT_2^D =& \eps_2 \left[\bE_0\bE^D_2 + \bE^D_2\bE_0 + \bE_2^D\bE_2^D\right.\\
	 &\left. -\frac{1}{2}\left(2\bE_0\cdot\bE^D_2+\bE^D_2\cdot\bE^D_2\right) \right].
	\nonumber
\end{aligned}
\end{equation}
Here $\bE_0$ is the applied electric field and $\bE^D_2$ is the disturbance field due to the particle. The detailed calculation process can be found in Appendix C. For both normal and tangential applied field, the force has only component in the direction normal to the wall surface (in the $z$-direction). 

For a normal electric field, the force on the particle is
\begin{equation} \label{FzResultN}
	F_z = -4 \eps_2 \pi \sum_{n=0}^{\infty} \left[ \lambda_n A_n^2 - (n+1) A_n A_{n+1} \right].
\end{equation}
 
For a tangential electric field, the force on the particle is
\begin{equation} \label{FzResultT}
	F_z = \frac{1}{2} \eps_2 \pi \left(I_1^T+I_2^T\right),
\end{equation}
where
\begin{widetext}
	\begin{align}
		I_1^T =& \sum_{n=1}^{\infty} \frac{A_n^2}{\lambda_n}n(n+1) -3\sum_{n=1}^{\infty} \frac{A_nA_{n+1}}{\lambda_n\lambda_{n+1}}n(n+1)(n+2) + \sum_{n=1}^{\infty}\sum_{m=1}^{\infty} A_nA_m \frac{n^2m^2}{\lambda_n\lambda_m} I_{n+1,m+1}	\nonumber \\
		& + \sum_{n=1}^{\infty}\sum_{m=1}^{\infty} A_nA_m \frac{(n+1)^2(m+1)^2}{\lambda_n\lambda_m} I_{n-1,m-1}
		- 2\sum_{n=1}^{\infty}\sum_{m=1}^{\infty} A_nA_m \frac{(n+1)^2m^2}{\lambda_n\lambda_m} I_{n-1,m+1},
		\label{I1T} \\
		I_2^T =& 4 \sum_{n=1}^{\infty}\sum_{m=1}^{\infty} A_nA_m I_{n,m}. \label{I2T}
	\end{align}
\end{widetext}
$I_{n,m}$ stands for the following integral
\begin{equation}
	I_{n,m} = \int_{-1}^{1} \frac{P_n^1(u)P_m^1(u)}{1+u} du
	= \begin{cases}
		n(n+1) & n=m\\
		(-1)^{p-q} q(q+1) & n \neq m
	\end{cases}
	,\nonumber
\end{equation}
where $p=\max(n,m)$ and $q=\min(n,m)$.

\subsection{Special case: perfectly conducting sphere in normal electric field}
A perfectly conducting sphere occurs in the limit of $\sigma_1 \to \infty$. This requires a special solution because $\beta_{12} \to 1$ and causes the denominator in \refeq{AnAsymptoticN} to become zero. Here we outline this special solution.

A perfectly conducting sphere implies a constant potential $\Phi_p$ along the interface. In this case, boundary condition \refeq{bc2} are replaced by $\Phi_2 = \Phi_p$. Here $\Phi_p$ is an unknown which we will determine using the zero net charge condition.

The solution approach for $\Phi_2$ parallels the approach at the beginning of this section. \refeq{gsN} still represents the solution $\hat{\Phi}_2$. Using the constant potential boundary condition, it is straightforward to show that
$$
\hat{\Phi}_2 \sqrt{h} \sum_{n=0}^{\infty} B_n \sinh\left(\lambda_n\eta\right) P_n(\cos\xi),
$$
where
$$
B_n = \sqrt{2} \left( \Phi_p +2 E_0 a \lambda_{n} \right) \frac{\exp (- \lambda_{n} \eta_{s})}{\sinh (\lambda_{n} \eta_{s})}.
$$
Now, paralleling the charge calculation in Appendix C, we find the net charge on the particle,
$$
Q_{net} = 4\pi \varepsilon_2 a \Phi_p S_1\left( \eta_s \right) + 8\pi \varepsilon_2 a^2 E_0 S_2 \left( \eta_s \right) = 0,
$$
where $S_1$ and $S_2$ are two sums
\begin{align}
	S_1\left( \eta_s \right) &= \sum_{n=0}^{\infty} \frac{\exp\left(-\lambda_n \eta_s\right)}{\sinh\left(\lambda_n \eta_s\right)}, \nonumber \\
	S_2\left( \eta_s \right) &= \sum_{n=0}^{\infty} \frac{\lambda_n \exp\left(-\lambda_n \eta_s\right)}{\sinh\left(\lambda_n \eta_s\right)}. \nonumber
\end{align}
Setting $Q_{net}$ to zero we can solve for $\Phi_p$ to find
$$
\Phi_p = -2E_0 a \frac{S_2 \left( \eta_s \right)}{S_1 \left( \eta_s \right)}.
$$
From \refeq{Force} with $\bT_1 = 0$,we can calculate the vertical force on the particle to find
\begin{equation} \label{SingularityInteractionForce}
	F_z = -\eps_2 \pi \sum_{n=0}^{\infty} \left[ \lambda_n B_n^2 - (n+1) B_n B_{n+1} \right].
\end{equation}

\subsection{Variation of the force with separation and media electric properties}
\begin{figure}[H]
	\centering
	\subfigure[Normal electric field]{\includegraphics[width=1\linewidth]{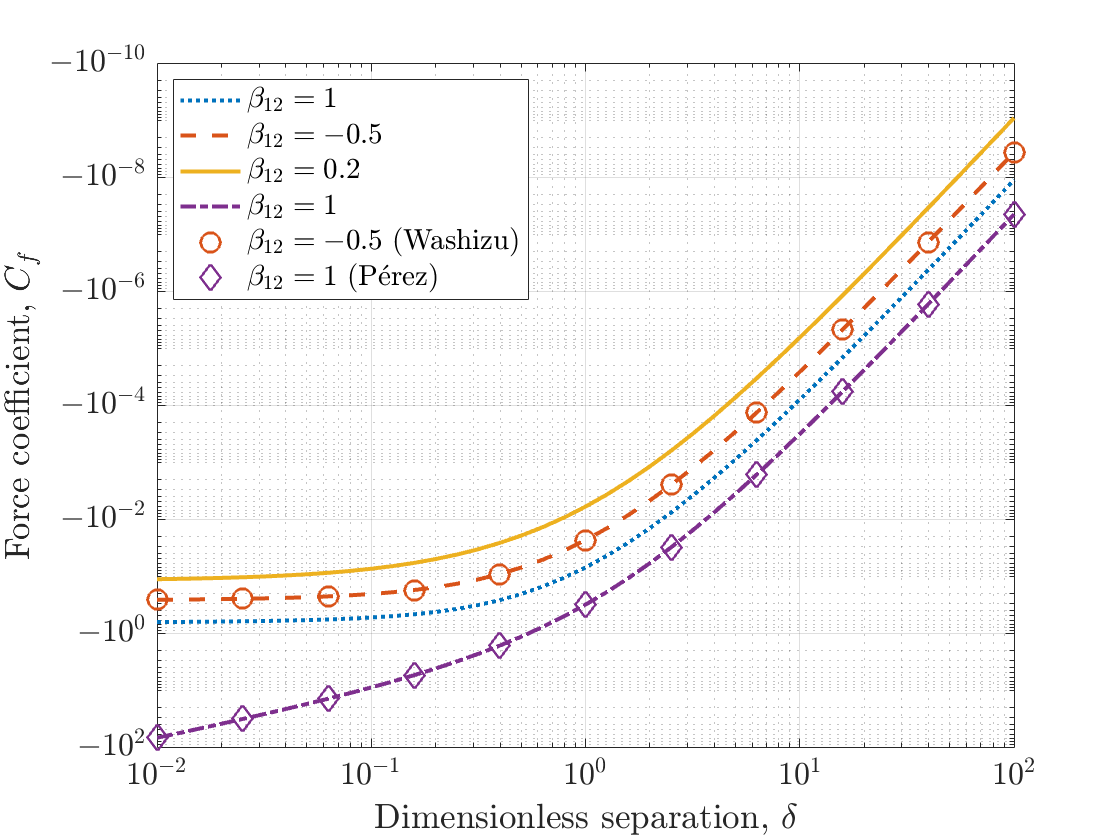}}
	\subfigure[Tangential electric field]{\includegraphics[width=1\linewidth]{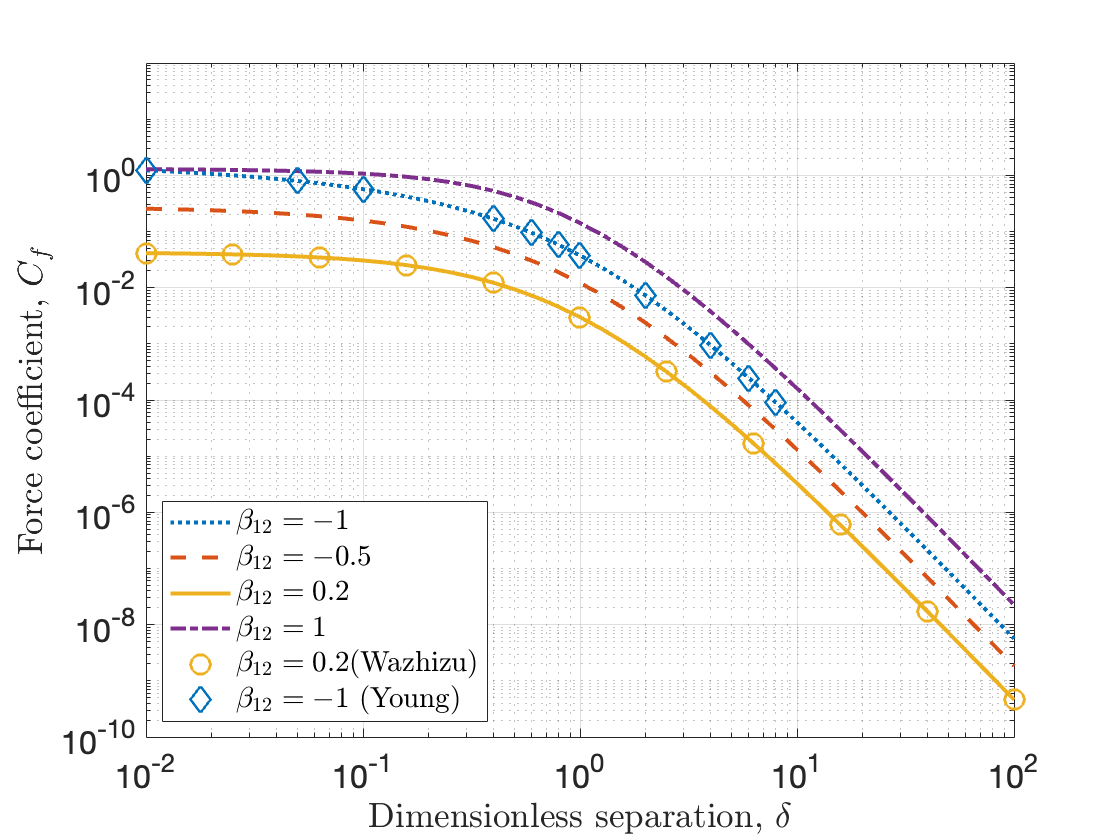}}
	\caption{Force coefficient $C_f =  {F_z}/{(\varepsilon_2 E_0^2 R^2)}$ as a function of dimensionless separation $\delta=s/R$. 
	Circles: Washizu's solution using the equivalent multipole method \cite{Washizu:1996}.
	Diamonds in (a): P{\'e}erez solution using the method of images for a conducting particle \cite{Perez:2002}
	Diamonds in (b): Young's solution using bispherical coordinates for a insulating particle \cite{young2005dielectrophoretic}
		}
	\label{fig:cfvsdelta}
\end{figure}

In the absence of a wall, the force on a charge-free particle in a uniform electric field is zero. The wall breaks the symmetry and, as shown below, gives rise to attraction in the case of a normal applied field and repulsion in the case of a tangential applied field. In \reffig{fig:cfvsdelta}, we plot the force coefficient $C_f = F_z/\left(\eps_2E_0^2R^2\right)$ as a function of particle-wall separation $\delta$. In both cases, the magnitude of the force increases  with decreasing separation.

\begin{figure}[H]
	\centering
	\subfigure[Normal electric field]{\includegraphics[width=1\linewidth]{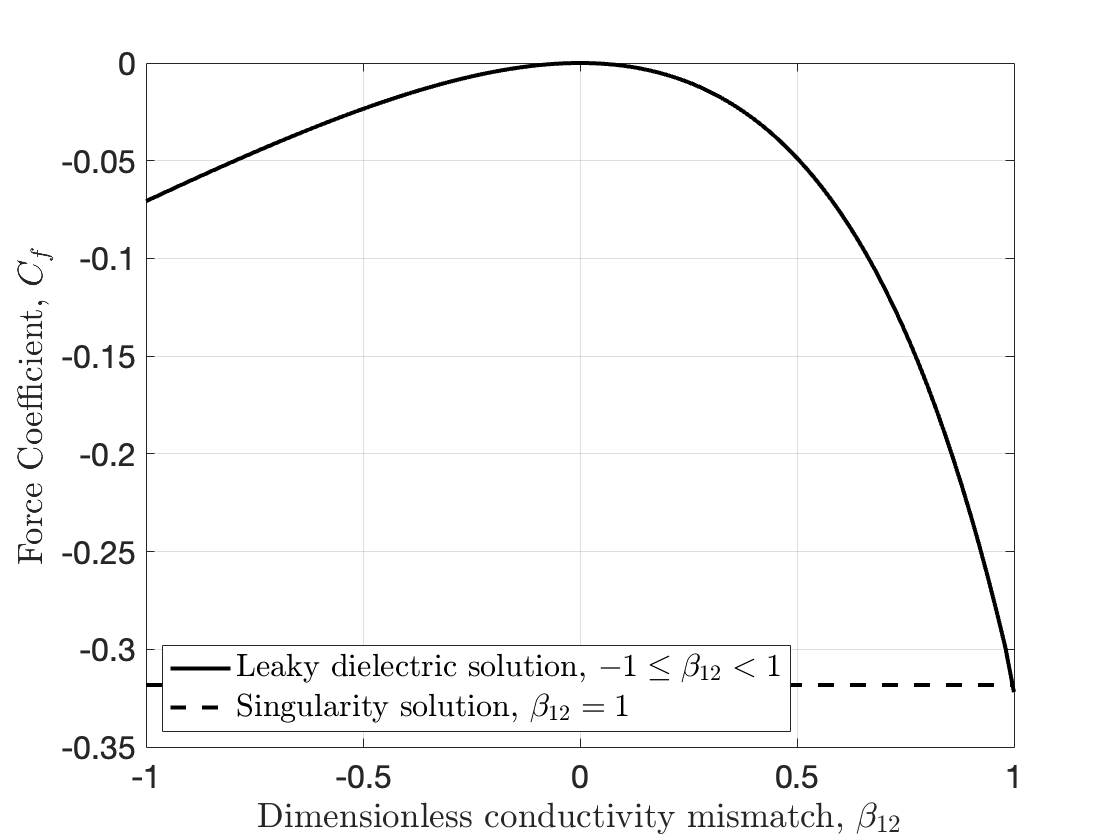}}
	\subfigure[Tangential electric field]{\includegraphics[width=1\linewidth]{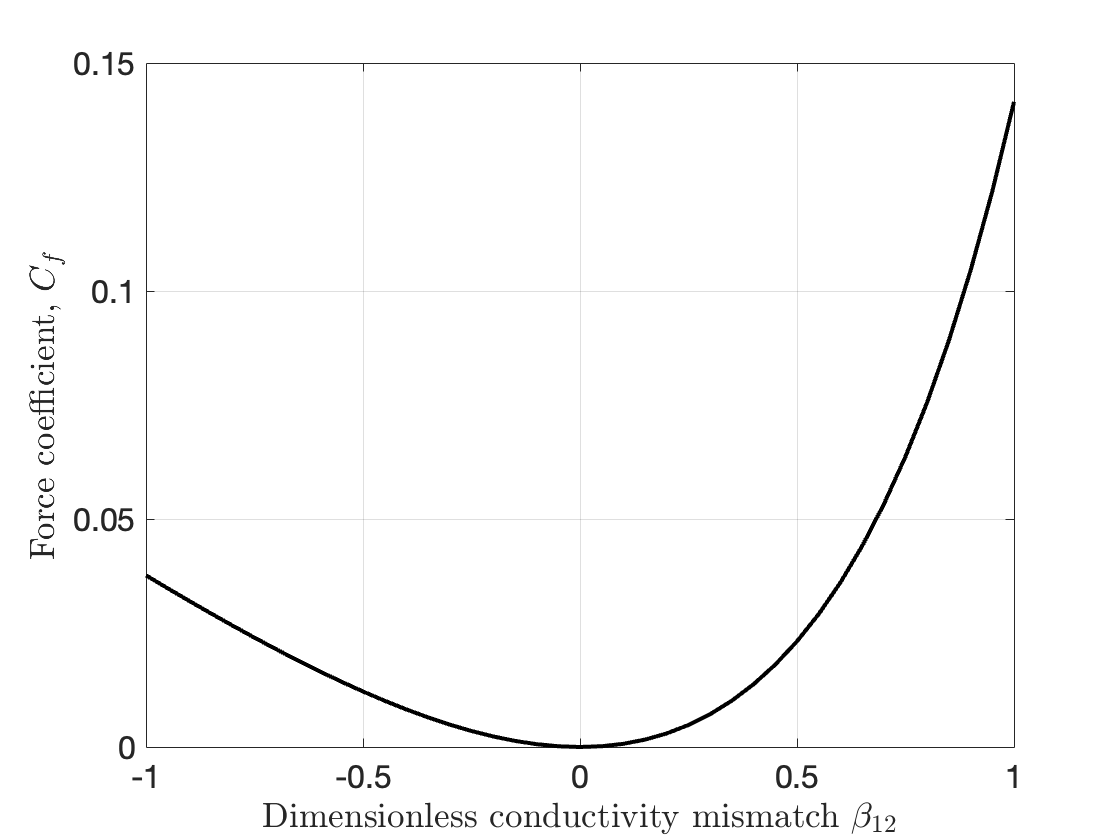}}
	\caption{Force coefficient as a function of $\beta_{12}$ (separation $s/R=1$). The dashed line corresponds to the perfectly conducting solution $\left(\beta_{12} = 1\right)$. }
	\label{fig:cfvsbeta}
\end{figure}

The dependence on the dimesionless conductivity mismatch $\beta_{12}$, however, is nonmonotonic. \reffig{fig:cfvsbeta} shows that the force vanishes for $\beta_{12}=0$, as expected since the particle and suspending media have the same electric properties. The variation with $\beta_{12}$ is asymmetric with respect to $\beta_{12}=0$. 
In the case of a normal applied field, a conducting particle is attracted more strongly by the electrode compared to an  insulating particle. Similar trend is observed also in the case of a tangential applied field: the conducting particle experiences stronger repulsion compared to the insulating particle.

\subsection{Asymptotic behavior of the force at large particle-wall separations} 
In this section, we present the asymptotic behavior of the force coefficient $C_f$ for large separations. In the case of normal electric field, the numerical solution for $A_n$ suggests a two-term approximation. For $\delta=s/R \gg 1$, from \refeq{eta_s}, $\eta_s \sim \ln(2\delta)$. Applying this result in \ref{LnN} and \ref{GnN}, we find the leading-order asymptotic form of the system \refeq{ThreeDiagonal} 
\begin{align}
	&\sqrt{2} \left(\beta_{12} -1\right) A_0 + 2\sqrt{2} A_1 \sim E_0 \beta_{12} R \frac{1}{\delta ^2} \label{AsymptoticEquation1} \\
	&\sqrt{2} A_0 + 2\sqrt{2} \left(\beta_{12} -3\right) \delta^2 A_1 \sim -2 E_0 \beta_{12} R
	\label{AsymptoticEquation2}
\end{align}

The asymptotic behavior of coefficients $A_0$ and $A_1$ is found from these two equations by balancing the  $O(\delta)$ terms,
\begin{equation} \label{AnAsymptoticLargeDelta}
	A_0 \sim -A_1 \sim -\frac{E_0 \beta_{12} R}{\sqrt{2} \left(3-\beta_{12}\right)} \frac{1}{\delta ^2}
\end{equation}
Substituting into \refeq{FzResultN} yields the asymptotic behavior of the force coefficient $C_f$ for the normal electric field,
\begin{equation} \label{CfAsymptoticN}
	C_f \sim - \frac{6\pi \beta_{12}^{2}}{\left(3-\beta_{12}\right)^2} \frac{1}{\delta ^4}
\end{equation}
In the case of a tangential applied electric field, the numerical solution for $A_n$ suggests a one-term approximation. The asymptotic form of \refeq{ThreeDiagonal} is
\begin{equation}
	\sqrt{2} \left(\beta_{12}-3\right)\delta^{5/2} A_1 \sim E_0\beta_{12}R \delta^{1/2},
\end{equation}
from which we find the asymptotic behavior of $A_1$,
\begin{equation}
	A_1 \sim \frac{E_0\beta_{12}R}{\sqrt{2}\left(\beta_{12}-3\right)} \frac{1}{\delta^2}
\end{equation}
Substituting into \refeq{FzResultT} yields the asymptotic behavior of the force coefficient $C_f$ for the tangential electric field
\begin{equation} \label{CfAsymptoticT}
	C_f \sim \frac{3\pi\beta_{12}^2}{\left(\beta_{12}-3\right)^2} \frac{1}{\delta^4}.
\end{equation}
To a leading order, the effect of the wall is equivalent  to placing a mirror image dipole.  The  $1/\delta^4$ dependence is consistent with the  dipole-dipole interaction \cite{yariv:2006}.

\begin{figure}[H]
	\centering
	\subfigure[Normal electric field]{\includegraphics[width=1\linewidth]{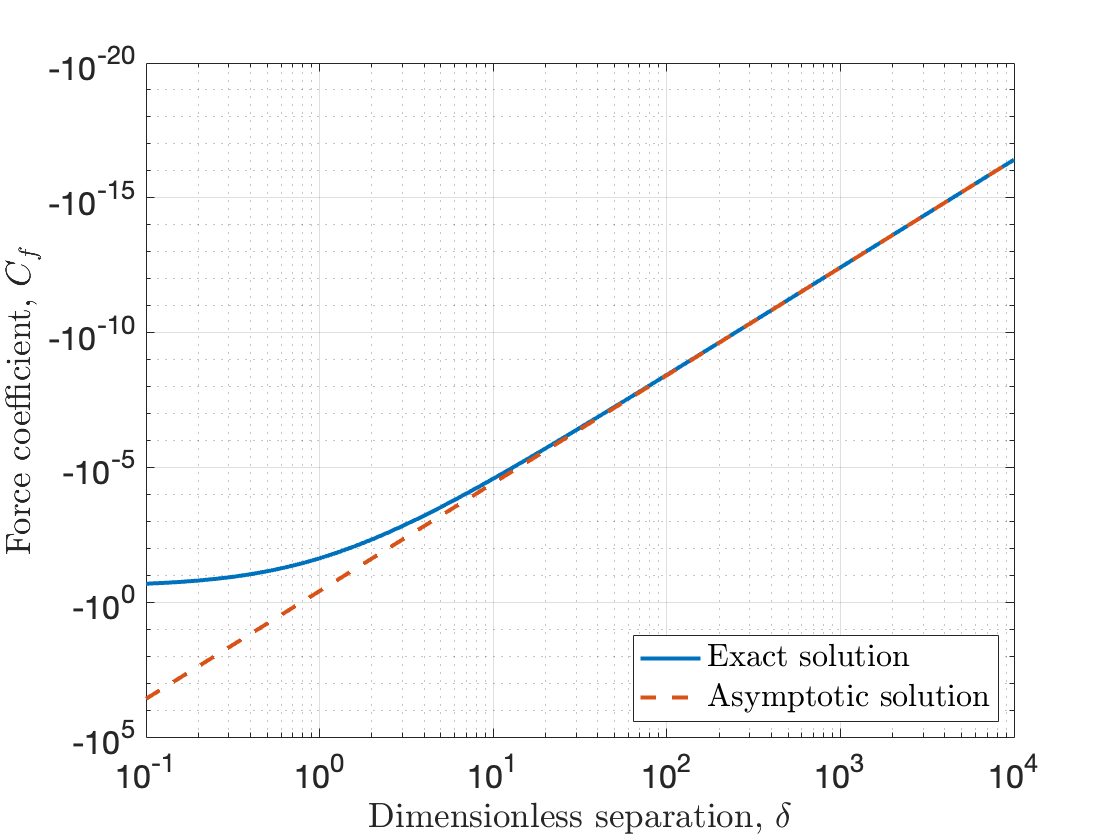}}
	\subfigure[Tangential electric field]{\includegraphics[width=1\linewidth]{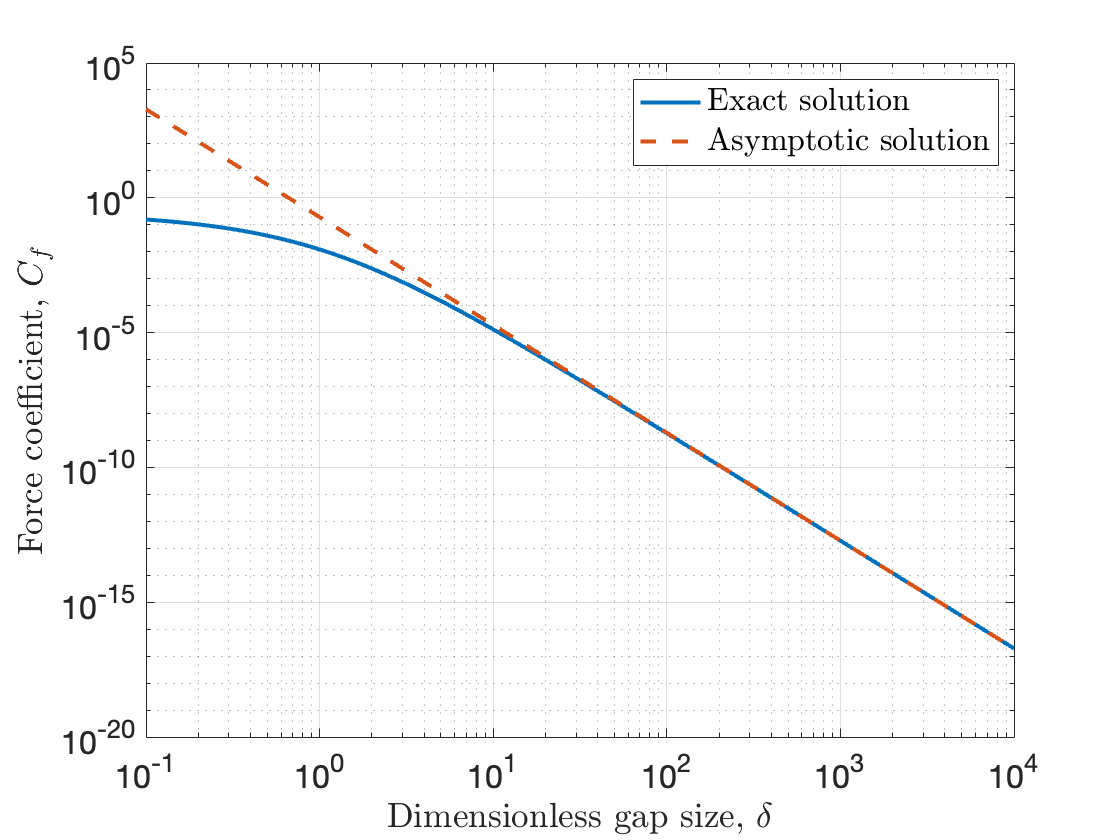}}
	\caption{Force coefficient asymptotic ($\beta_{12}=-0.5$)
	}
	\label{fig:cfasymptotic}
\end{figure}

The exact  and asymptotic solutions are compared  in \reffig{fig:cfasymptotic}. The agreement is excellent when $\delta > 10$. 
\refeq{CfAsymptoticN} is also valid for the special case $\beta_{12}=1$.



\section{Conclusions}
In this paper, the interaction of a  spherical particle and a planar wall  in the presence of a uniform electric field is studied in the framework of the leaky dielectric model. The bulk media are assumed to be charge free; charge brought by conduction is only present at the surfaces separating media with different conductivities and permittivities.  Analytical solutions of the Laplace equation for the electric potential are derived using separation of variables in bispherical coordinates. The force  on the particle is calculated using the Maxwell stress tensor.  We find that in the case of normal electric field, which corresponds to a particle near an electrode, the force is always attractive but at a given separation it varies nonmonotonically  with particle-suspending medium conductivity ratio; the force on a particle that is more conducting than the suspending medium is much larger compared to the force on a particle less conducing than the suspending medium. In the case of tangential electric field, which  corresponds to a particle near an insulating boundary, the force is always repulsive. 
In both cases, the force decreases with increasing separation between the particle and the wall and its asymptotic behavior at large separation is given by the force between the particle dipole and its mirror image. The results provide a comprehensive understanding of the effect of particle conductivity on the wall-induced electrostatic force.

\section{Acknowledgements}

This research was supported by NSF awards DMS-2108502 and DMR-2004926.

\appendix
\section{Components of the electric field strength}

In the case of the normal electric field, the components of the electric field strength are given by
\begin{widetext}
\begin{align}
	E_{1,\eta} &= \frac{E_0}{h}\left(1-\cosh\eta\cos\xi\right) - \frac{\sqrt{h}\sinh\eta}{2a} \sum_{n=0}^{\infty} A_n \left[e^{2\lambda_n\eta_s}-1\right]e^{-\lambda_n\eta}P_n(\cos\xi)   + \frac{h^{3/2}}{a} \sum_{n=0}^{\infty} A_n\lambda_n\left[e^{2\lambda_n\eta_s}-1\right]e^{-\lambda_n\eta}P_n(\cos\xi) \\
	E_{1,\xi} =& -\frac{E_0}{h} \sinh\eta\sin\xi - \frac{\sqrt{h}\sin\xi}{2a} \sum_{n=0}^{\infty} A_n\left[e^{2\lambda_n\eta_s}-1\right]e^{-\lambda_n\eta}P_n(\cos\xi)  - \frac{h^{3/2}}{a} \sum_{n=0}^{\infty} A_n\left[e^{2\lambda_n\eta_s}-1\right]e^{-\lambda_n\eta})\frac{d}{d\xi}P_n(\cos\xi) \\
	E_{2,\eta} =& \frac{E_0}{h}\left(1-\cosh\eta\cos\xi\right) - \frac{\sqrt{h}\sinh\eta}{a}\sum_{n=0}^{\infty} A_n \sinh\left(\lambda_n\eta\right)P_n(\cos\xi)  - \frac{2h^{3/2}}{a} \sum_{n=0}^{\infty} A_n\lambda_n \cosh\left(\lambda_n\eta\right)P_n(\cos\xi) \\
	E_{2,\xi} =& -\frac{E_0}{h} \sinh\eta\sin\xi - \frac{\sqrt{h}\sin\xi}{a} \sum_{n=0}^{\infty} A_n\sinh\left(\lambda_n\eta\right)P_n(\cos\xi) -\frac{2h^{3/2}}{a} \sum_{n=0}^{\infty} A_n\sinh\left(\lambda_n\eta\right)\frac{d}{d\xi}P_n(\cos\xi) 
\end{align}
\end{widetext}

$E_{1,\phi}$ and $E_{2,\phi}$ are both zero due to the symmetry.

In the case of the tangential electric field, the components of the electric field strength are given by
\begin{widetext}
\begin{align}
	E_{i,\eta} &= \hat{E}_{i,\eta} \sin\phi \,,E_{i,\xi} = \hat{E}_{i,\xi} \sin\phi \,,E_{i,\phi} = \hat{E}_{i,\phi} \cos\phi. \nonumber
\end{align}
\begin{align}
	\hat{E}_{1,\eta} =& -\frac{E_0}{h}\sinh\eta\sin\xi - \frac{\sqrt{h}\sinh\eta}{2a} \sum_{n=1}^{\infty} A_n \left[1+e^{2\lambda_n\eta_s}\right] e^{-\lambda_n\eta} P_n^1(\cos\xi) + \frac{h^{3/2}}{a} \sum_{n=1}^{\infty} A_n\lambda_n \left[1+e^{2\lambda_n\eta_s}\right] e^{-\lambda_n\eta} P_n^1(\cos\xi),
	\label{HatE1eta} \\
		\hat{E}_{1,\xi} =& \frac{E_0}{h}\left(\cosh\eta\cos\xi-1\right) - \frac{\sqrt{h}\sin\xi}{2a}\sum_{n=1}^{\infty} A_n \left[1+e^{2\lambda_n\eta_s}\right]e^{-\lambda_n\eta}  P_n^1(\cos\xi) 
	- \frac{h^{3/2}}{a} \sum_{n=1}^{\infty} A_n \left[1+e^{2\lambda_n\eta_s}\right] e^{-\lambda_n\eta} \frac{d}{d\xi}P_n^1(\cos\xi),
	\label{HatE1xi} \\
	\hat{E}_{1,\phi} =& E_0 - \frac{h^{3/2}}{a\sin\xi} \sum_{n=1}^{\infty} A_n \left[1+e^{2\lambda_n\eta_s}\right] e^{-\lambda_n\eta} P_n^1(\cos\xi),
	\label{HatE1phi} \\
		\hat{E}_{2,\eta} =& -\frac{E_0}{h}\sinh\eta\sin\xi - \frac{\sqrt{h}\sinh\eta}{a} \sum_{n=1}^{\infty} A_n \cosh\left(\lambda_n\eta\right) P_n^1(\cos\xi) 
	-\frac{2h^{3/2}}{a} \sum_{n=1}^{\infty} A_n\lambda_n \sinh\left(\lambda_n\eta\right)P_n^1(\cos\xi),
	\label{HatE2eta} \\
		\hat{E}_{2,\xi} =& \frac{E_0}{h}\left(\cosh\eta\cos\xi-1\right) - \frac{\sqrt{h}\sin\xi}{a}\sum_{n=1}^{\infty} A_n \cosh\left(\lambda_n\eta\right)P_n^1(\cos\xi) - \frac{2h^{3/2}}{a} \sum_{n=1}^{\infty} A_n \cosh\left(\lambda_n\eta\right)\frac{d}{d\xi}P_n^1(\cos\xi),
	\label{HatE2xi} \\
		\hat{E}_{2,\phi} =& E_0 - \frac{2h^{3/2}}{a\sin\xi} \sum_{n=1}^{\infty} A_n\cosh\left(\lambda_n\eta\right)P_n(\cos\xi).
	\label{HatE2phi}
	\end{align}
\end{widetext}
\section{Recurrence Formula for $A_n$}
Here we outline the derivation of the recurrence formula which is used to determine coefficients $A_n$ in the expression of the electric potential.
\subsection{Normal electric field}
For the normal applied electric field, the disturbance fields $\hat{\Phi}_{1}$ and $\hat{\Phi}_{2}$ can be written as
\begin{equation}
	\hat{\Phi}_{1}=\sqrt{h}\tilde{\Phi}_{1},\hat{\Phi}_{2}=\sqrt{h}\tilde{\Phi}_{2}.\label{DisturbaceField_V}
\end{equation}
where
\begin{align}
	\tilde{\Phi}_{1}&=\sum_{n=0}^{\infty}A_{n}\left[e^{2\lambda_n\eta_s}-1\right]
	e^{-\lambda_{n}\eta}P_{n}\left(\cos\xi\right)\label{DisturbaceField_V1}\\
	\tilde{\Phi}_{2}&=\sum_{n=0}^{\infty}2A_{n}\sinh\left(\lambda_{n}\eta\right)P_{n}\left(\cos\xi\right)\label{DisturbaceField_V2}
\end{align}
Substituting equation \refeq{DisturbaceField_V} into the boundary condition for the continuity of the normal electric current,
\[
\sigma_1 \frac{\partial \hat{\Phi}_{1}}{\partial \eta} - \sigma_2 \frac{\partial \hat{\Phi}_{2}}{\partial \eta} = E_0 \left(\sigma_1-\sigma_2\right) \frac{\partial z}{\partial \eta},
\]
leads to
\begin{equation}
\begin{split}
	2&\left(\cosh\eta_{s}-u\right)  \left(\frac{\sigma_{1}}{\sigma_{1}+\sigma_{2}} \frac{\partial\tilde{\Phi}_{1}}{\partial\eta} - \frac{\sigma_{2}}{\sigma_{1}+\sigma_{2}} \frac{\partial\tilde{\Phi}_{2}}{\partial\eta}\right) \\
	&+ \sinh\eta_{s}\beta_{12}\tilde{\Phi}_{2} = 2E_{0}\beta_{12} \left(\cosh\eta_{s}-u\right)^{1/2} \frac{\partial z}{\partial \eta}
	\label{recurrence_V1}
		\end{split}
\end{equation}
where $u=\cos\xi$. Plugging \refeq{DisturbaceField_V1} and \ref{DisturbaceField_V2} into the left hand side of \refeq{recurrence_V1}, one can gets
\begin{widetext}
\begin{align}
	{\rm L.H.S}=&\sum_{n=0}^{\infty}\left[\left( \beta_{12}\sinh\eta_{s} - 2\lambda_{n}\cosh\eta_{s} \right) \exp\left(\lambda_{n}\eta_{s}\right) 
	- \beta_{12}\left( \sinh\eta_{s} - 2\lambda_{n}\cosh\eta_{s} \right) \exp\left(-\lambda_{n}\eta_{s}\right)\right] A_{n}P_{n}(u) 
	\nonumber \\
	&+\sum_{n=0}^{\infty}A_{n}\left[\exp\left(\lambda_{n}\eta_{s}\right)-\beta_{12}\exp\left(-\lambda_{n}\eta_{s}\right)\right]2\lambda_{n}uP_{n}(u) \nonumber
\end{align}
\end{widetext}
Note that $\lambda_{n}\equiv n+1/2$ and $\beta_{12}=\left(\sigma_{1}-\sigma_{2}\right)/\left(\sigma_{1}+\sigma_{2}\right)$. With the help of the recurrence formula for Legendre polynomials,
\begin{equation}
	(2n+1)uP_{n}(u)=(n+1)P_{n+1}(u)+nP_{n-1}(u)\label{recurrence_Pn}
\end{equation}
the second sum of $L$ could be could be written as
\begin{widetext}
\begin{equation}
	\sum_{n=1}^{\infty} n \left[ \exp\left(\lambda_{n-1}\eta_{s}\right)-\beta_{12}\exp\left(-\lambda_{n-1}\eta_{s}\right) \right] A_{n-1}P_{n}(u) + \sum_{n=0}^{\infty} (n+1) \left[\exp\left(\lambda_{n+1}\eta_{s}\right)-\beta_{12} \exp\left(-\lambda_{n+1}\eta_{s}\right)\right] A_{n+1}P_{n}(u) \nonumber
\end{equation} 
\end{widetext}
Putting these terms together, we have
\begin{equation}
\begin{split}
	{\rm L.H.S}=& \sum_{n=1}^{\infty} L_{n,1}A_{n-1}P_{n}(u) + \sum_{n=0}^{\infty} L_{n,2}A_{n}P_{n}(u) \\
	&+ \sum_{n=0}^{\infty} L_{n,3}A_{n+1}P_{n}(u).
	\label{LHS_V}
	\end{split}
\end{equation}
Coefficients $L_{n,1}$, $L_{n,2}$, and $L_{n,3}$ are listed below,
\begin{equation}
\begin{split}
	L_{n,1} =& n \left( e^{\lambda_{n-1}\eta_{s}} - \beta_{12} e^{-\lambda_{n-1}\eta_{s}} \right), \\
	L_{n,2} =& \left( \beta_{12} \sinh\eta_{s} - 2\lambda_{n}\cosh\eta_{s} \right) e^{\lambda_{n}\eta_{s}}  \\
	& - \beta_{12} \left( \sinh\eta_{s} - 2\lambda_{n}\cosh\eta_{s} \right) e^{-\lambda_{n}\eta_{s}},  \\
	L_{n,3} =& (n+1) \left( e^{\lambda_{n+1}\eta_{s}} - \beta_{12} e^{-\lambda_{n+1}\eta_{s}} \right), 
	\end{split}
 \label{LnN}
\end{equation}
The next step is to expand the right hand side of \refeq{recurrence_V1} in terms of Legendre polynomials. Note that
$$
z = \frac{a}{h} \sinh\eta.
$$
Substituting it into the right hand side of \refeq{recurrence_V1}, we have
\begin{equation}
	{\rm R.H.S}=2E_{0}a\beta_{12}\frac{1-u\cosh\eta_{s}}{h_s^{3/2}}, \label{RHS_V1}
\end{equation}
where $h_s = \cosh \eta_s - u$. The generating function of Legendre polynomials is 
\begin{equation}
	\frac{1}{\left(1-2ut+t^2\right)^{1/2}}=\sum_{n=0}^{\infty}t^{n}P_{n}(u) \label{Pn_GeneratingFun}
\end{equation}
Letting $t = \exp\left(-\eta_s\right)$ yields
\begin{equation} \label{LegendreExpansion1}	
	\frac{1}{h_s^{1/2}}=\sqrt{2}\sum_{n=0}^{\infty}\exp\left(-\lambda_{n}\eta_{s}\right)P_{n}(u)
\end{equation}
Calculating derivative with respect to $\eta_{s}$, we get
\begin{equation} 
	\frac{1}{h_s^{3/2}}=\frac{2\sqrt{2}}{\sinh\eta_{s}}\sum_{n=0}^{\infty}\lambda_{n}\exp\left(-\lambda_{n}\eta_{s}\right)P_{n}(u)
	\label{RHS_V2}
\end{equation}
Multiplying \refeq{RHS_V2} by $u$ and using the recurrence formula \refeq{recurrence_Pn}, we obtain
\begin{equation} \label{RHS_V3}
	\frac{u}{h_s^{3/2}} = \frac{\sqrt{2}}{\sinh\eta_{s}} \sum_{n=0}^{\infty} \left[n e^{-\lambda_{n-1}\eta_{s}} + (n+1)e^{-\lambda_{n+1}\eta_{s}}\right]P_{n}(u)
\end{equation}
Using \refeq{RHS_V2} and \ref{RHS_V3}, we could expand \refeq{RHS_V1} as
\begin{equation} \label{RHS_V}
	{\rm R.H.S} = \sum_{n=0}^{\infty} G_n P_{n}(u), 
\end{equation}
where $G_n$ is
\begin{equation} \label{GnN}
\begin{split}
	G_n =& -2\sqrt{2}E_0a\beta_{12} \left\{ \frac{\cosh\eta_{s}}{\sinh\eta_{s}} \left[ ne^{-\lambda_{n-1}\eta_{s}} + (n+1)e^{-\lambda_{n+1}\eta_{s} }\right]\right. \\
	&\left. - \frac{2}{\sinh\eta_{s}}\lambda_{n} e^{-\lambda_{n}\eta_{s}}\right\}
	\end{split}
\end{equation}
Equating \refeq{LHS_V} and \ref{RHS_V} leads to the recurrence formula we need.

For $n \gg 1$, \refeq{ThreeDiagonal} has the following asymptotic form
\begin{widetext}
\begin{equation}
	-2n t^2 A_{n-1} + \left[ (2n+1) (1+t^2) - \beta_{12} (1-t^2)\right] A_n - 2(n+1) A_{n+1} \sim 4\sqrt{2} \frac{E_0 a \beta_{12}}{1-t^2} t^{2n+1} \left[ n-(2n+1)t^{2} + (n+1) t^4\right], \nonumber
\end{equation}
\end{widetext}
where $t=\exp(-\eta_{s})$. It gives the asymptotic behavior of coefficients $A_n$ for large $n \gg 1$,
\begin{equation} \label{AnAsymptoticN}
	A_n \sim 4\sqrt{2} \frac{E_0 a \beta_{12}}{1-t^2} t^{2n+1} \left[\frac{n}{3-\beta_{12}} (1-t^2) - \frac{1}{1-\beta_{12}} t^2\right].
\end{equation}

\subsection{Tangential electric field}
For the tangential electric field, the recurrence formula could be obtained following similar steps. Results are listed below
\begin{equation} \label{LnT}
\begin{split}
	L_{n,1} =& \frac{n-1}{2} \left( e^{\lambda_{n-1}\eta_s} + \beta_{12} e^{-\lambda_{n-1}\eta_s} \right),  \\
	L_{n,2} =& \left( \frac{\beta_{12}}{2} \sinh \eta_s - \lambda_n \cosh \eta_s \right) e^{\lambda_n \eta_s}  \\
	&+ \beta_{12} \left( \frac{1}{2} \sinh\eta_s - \lambda_n \cosh\eta_s \right) e^{-\lambda_n \eta_s}, \\
	L_{n,3} =& \frac{n+2}{2} \left( e^{\lambda_{n+1}\eta_s} + \beta_{12} e^{-\lambda_{n+1}\eta_s} \right), \\
		\end{split}
\end{equation} 
\begin{equation} \label{GnT}
\begin{split}	
	G_n =& 2\sqrt{2} E_0 a \beta_{12} \sinh\eta_s e^{-\lambda_n\eta_s},
	\end{split}
\end{equation} 
where $n\ge 1$. For $n \gg 1$,
\begin{equation}
	A_n \sim \frac{4\sqrt{2}E_0 a \beta_{12}}{\beta_{12}-3} t^{2n+1}. \label{AnAsymptoticT}
\end{equation}

\section{Electric force}

Consider the case of a normal applied field $\bE_0 = E_0 \hat{\bf z}$. At $\eta=0$, the disturbance field is given by
\begin{align}
	&E^D_{2,\eta} = - \frac{h_0^{3/2}}{a} \sum_{n=0}^{\infty} 2 A_n \lambda_n P_n\left( \cos\xi \right) \label{E2D_eta} \\
	&E^D_{2,\xi} = E^D_{2,\phi} = 0,
\end{align}
where $h_0 = 1-\cos\xi$. Due to the symmetry of this problem, the force is in $z$-direction,
\begin{align}
	F_z =& -\int_{S_B} \mathbf{e}_z \cdot \mathbf{T}_2^D \cdot \mathbf{n} dS         \nonumber \\
	=& -\varepsilon_2 \int_{S_B} E_0E_{2,\eta}^D + \frac{1}{2}
	\left(E_{2,\eta}^D\right)^2 dS \nonumber \\
	=& -\varepsilon_2 \pi a^2 \int_{0}^{\pi} \frac{\sin\xi}{h_0^2}  \left[ 2E_0E_{2,\eta}^D + \left(E_{2,\eta}^D\right)^2 \right] d\xi \nonumber \\
	=& \varepsilon_2 \pi a^2 \left(I^N_1-I^N_2\right),  \label{FzV}
\end{align}
where $I^N_1$ and $I^N_2$ are two integrals
\begin{align}
	I^N_1 &= -2 \int_{0}^{\pi} \frac{\sin\xi}{h_0^2} E_0 E^D_{2,\eta} d\xi \label{I1V_def} \\
	I^N_2 &= \int_{0}^{\pi} \frac{\sin\xi}{h_0^2} \left( E^D_{2,\eta} \right)^2 d\xi \label{I2V_def}
\end{align}
Substituting \refeq{E2D_eta} into \refeq{I1V_def}, we have
\begin{align}
	I^N_1 =& \int_{-1}^{1} \frac{2E_0}{a\sqrt{1-u}} \sum_{n=0}^{\infty} 2A_n \lambda_n P_n(u) du \nonumber \\
	=& \frac{4E_0}{a} \sum_{n=0}^{\infty} A_n \lambda_n \int_{-1}^{1} \frac{P_n(u)}{\sqrt{1-u}} du. \nonumber
\end{align}
From the generating function \refeq{Pn_GeneratingFun}, we have
\begin{equation}
	\int_{-1}^{1} \frac{1}{\left(1-2ut+t^2\right)^{1/2}} P_n du = \sum_{m=0}^{\infty} t^m \int_{-1}^{1} P_m(u)P_n(u) du = \frac{t^n}{\lambda_n}. \nonumber
\end{equation}
Here the orthogonality of Legendre polynomials is used,
\begin{equation}
	\int_{-1}^{1} P_m(u)P_n(u) du = \frac{2}{2n+1} \delta_{n,m} = \frac{\delta_{n,m}}{\lambda_n}. \label{Pn_Orthogonality}
\end{equation}
Letting $t=1$ yields
\begin{equation}
	\int_{-1}^{1} \frac{P_n(u)}{\sqrt{1-u}} du = \frac{\sqrt{2}}{\lambda_n}. \nonumber	
\end{equation}
Substituting it into $I^N_1$, we get
\begin{equation}
	I_1^N = 4\sqrt{2}\frac{E_0}{a} \sum_{n=0}^{\infty} A_n \label{I1V_Result}
\end{equation}

For $I^N_2$, substituting \refeq{E2D_eta} into \refeq{I2V_def} yields
\begin{align}
	I^N_2 =& \int_{-1}^{1} \frac{1-u}{a^2} \left(\sum_{n=0}^{\infty} 2 A_n \lambda_n P_n(u)\right)^2 du \nonumber \\
	=& \sum_{m=0}^{\infty} \sum_{n=0}^{\infty} \frac{4}{a^2} A_m A_n \lambda_m \lambda_{n} \int_{-1}^{1} \left(1-u\right) P_n\left(u\right) P_m\left(u\right) du \nonumber	
\end{align}
Using the recurrence formula \refeq{recurrence_Pn} and orthogonality \refeq{Pn_Orthogonality}, the integral in $I^N_2$ could be calculated explicitly,
\begin{equation}
\begin{split}
	&\int_{-1}^{1} (1-u) P_n(u)P_m(u) du =\\
	& \int_{-1}^{1} P_n(u)P_m(u) du - \frac{n+1}{2n+1} \int_{-1}^{1} P_{n+1}(u)P_m(u) du  \\
	&- \frac{n}{2n+1} \int_{-1}^{1} P_{n-1}(u)P_m(u) du \nonumber \\
	&= \frac{\delta_{n,m}}{\lambda_n} - \frac{n+1}{2n+1} \frac{\delta_{n,m+1}}{\lambda_m} - \frac{n}{2n+1} \frac{\delta_{n,m-1}}{\lambda_m} 
\end{split}
\label{I2V_Process}
\end{equation}
Plugging \refeq{I2V_Process} into $I_2^N$, we have
\begin{equation}
	I_2^N = \frac{4}{a^2} \sum_{n=0}^{\infty} \left[ \lambda_nA_n^2 - (n+1)A_nA_{n+1} \right]. \label{I2V_Result}
\end{equation}
From \refeq{FzV}, we could get the solution for the electric force. A similar calculation can be done for a tangential electric field.

It is interesting to note that because $\bE$ is divergence free, plus the boundary condition \ref{bc2}, the net charge on the particle is
$$
Q_{net} = \oint_{S_p} \left(\eps_1\bE_1 - \eps_2\bE_2\right) \cdot {\bf e}_{\eta} dS = 0. 
$$
Also since $\oint_{S_p}\bE_2\cdot{\bf e}_{\eta}dS = \int_{S_B}\bE_2^D\cdot{\bf e}_{\eta} = 0$, we have from integrating \refeq{E2D_eta} that $\sum_{n=0}^{\infty} A_n = 0$. Hence $I_1^N = 0$.

\newpage
\bibliographystyle{unsrt}
\bibliography{refs}

\end{document}